\documentstyle[12pt,axodraw,epsfig]{article}

\newlength{\dinwidth}
\newlength{\dinmargin}
\def\lapproxeq{\lower .7ex\hbox{$\;\stackrel{\textstyle
<}{\sim}\;$}}
\def\gapproxeq{\lower .7ex\hbox{$\;\stackrel{\textstyle
>}{\sim}\;$}}

\def\beq{\begin{equation}}
\def\eeq{\end{equation}}
\def\bea{\begin{eqnarray}}
\def\eea{\end{eqnarray}}

\def\GeV{\rm GeV}
\def\msb{\overline{\rm MS}}

\begin{document}
\titlepage
\begin{flushright}
Cavendish-HEP-2006/01 \\

\end{flushright}

\vspace*{0.5cm}

\begin{center}
{\Large \bf A Variable-Flavour Number Scheme for NNLO. }

\vspace*{1cm}
\textsc{R.S. Thorne\footnote{Royal 
Society University Research Fellow.}} \\

\vspace*{0.5cm} 
Cavendish Laboratory, University of Cambridge, \\ JJ Thomson Avenue,
Cambridge, CB3 0HE, UK
\end{center}

\vspace*{0.5cm}

\begin{abstract}
At NNLO it is particularly important to have a Variable-Flavour Number 
Scheme (VFNS) to deal with heavy quarks because there are major problems
with both the zero mass variable-flavour number scheme and the fixed-flavour 
number scheme. I illustrate these problems and    
present a general formulation of a Variable-Flavour Number Scheme (VFNS)
for heavy quarks that is explicitly 
implemented up to NNLO in the strong coupling 
constant $\alpha_S$, and may be 
used in NNLO global fits for parton distributions. The procedure combines
elements of the ACOT($\chi$) scheme and the Thorne-Roberts scheme.
Despite the fact that at NNLO the parton distributions are discontinuous 
as one changes the number of active quark flavours, all physical quantities 
are continuous at flavour transitions and the comparison with 
data is successful. 
\end{abstract}

\vspace*{0.3cm}
\section{Introduction}

While up, down and strange quarks may be treated as being effectively massless
partons, the heavy quarks, charm $\sim 1.5 \GeV$, bottom 
$\sim 4.3 \GeV$, top $\sim 175 \GeV$, must have their mass, $m_H$ 
taken into account 
in any QCD calculations. In particular it is essential to 
treat charm and bottom correctly in global fits for parton distributions.
There are two distinct regimes that can be considered.
Near threshold, $Q^2\sim m_H^2$, massive quarks are not treated as parton
constituents of the proton but are 
created in the final state. Any processes may be described using the 
Fixed-Flavour Number Scheme (FFNS). For example, 
structure functions are given by
\beq
F(x,Q^2)=C^{FF}_k(Q^2/m_H^2,Q^2/\mu^2)\otimes f^{n_f}_k(\mu^2),
\eeq
up to higher twist $({\cal O}(\Lambda^2/Q^2))$ corrections, where 
$n_f$ is the number of light partons and all the mass dependence is 
in the hard coefficient functions which have been calculated up to NLO
(i.e. ${\cal O}(\alpha_S^2)$) \cite{nlocalc}. 
This is reliable for scales not much 
greater than $m_H^2$, but increasing orders in $\alpha_S$ contain 
increasing logarithms in $Q^2/m_H^2$, and order-by-order perturbation theory 
is not guaranteed to be accurate. Also, the FFNS coefficient functions are
not known yet at NNLO, rendering an NNLO FFNS impossible to define. 

At high scales, i.e.  $Q^2, \mu^2 \gg m_H^2$, the heavy quarks are expected 
to behave like massless partons. The heavy quark is treated like the other 
partons and $\ln(\mu^2/m_H^2)$ terms are then 
automatically summed via evolution. The simplest approach is the 
Zero Mass Variable-Flavour Number Scheme (ZMVFNS) \cite{ZMVFNS}.
This ignores all ${\cal O}(m_H^2/Q^2)$ corrections for each of the 
$n_H$ heavy quarks , and the structure 
functions are given by
\beq
F(x,Q^2) = C^{ZMVF}_j(Q^2/\mu^2)\otimes f^{n_f+n_H}_j(\mu^2),
\eeq
where the hard coefficient functions are mass independent.  
Although this is 
called a ``scheme'' it is important to note that unlike usual scheme 
definitions, which are alternative ways to order the perturbative series, the
ZMVFNS is incorrect by terms of ${\cal O}(m_H^2/Q^2)$, and
is really only an approximation in the region $m_H^2\sim Q^2$. The 
approximation in this region may indeed be very important in practice,
as I will demonstrate later. A correct
variable-flavour number scheme should not have these inaccuracies, but should
correct the coefficient functions for the mass effects. 

As we go from an $n_f$-flavour to an $n_f+1$-flavour scheme, the
partons in the different number regions are related to each other 
perturbatively,
\beq
f^{n_f+1}_j(\mu^2)= A_{jk}(\mu^2/m_H^2)
\otimes f^{n_f}_k(\mu^2),
\eeq
where the perturbative matrix elements $A_{jk}(\mu^2/m_H^2)$
contain the $\ln(\mu^2/m_H^2)$ terms which  
relate $f^{n_f}_k(\mu^2)$ and $f^{n_f+1}_k(\mu^2)$ and lead to the  
correct evolution for both. There is then a similar relationship 
as we go from an $n_f+1$-flavour to an $n_f+2$-flavour scheme, 
but I will consider the transitions one at a time in this paper. At the
bottom quark transition point $n_f$ is effectively equal to 4, i.e. 
the charm quark is already evolving like a massless partons below this point.

\begin{figure}
\begin{center}
\centerline{\epsfxsize=0.42\textwidth\epsfbox{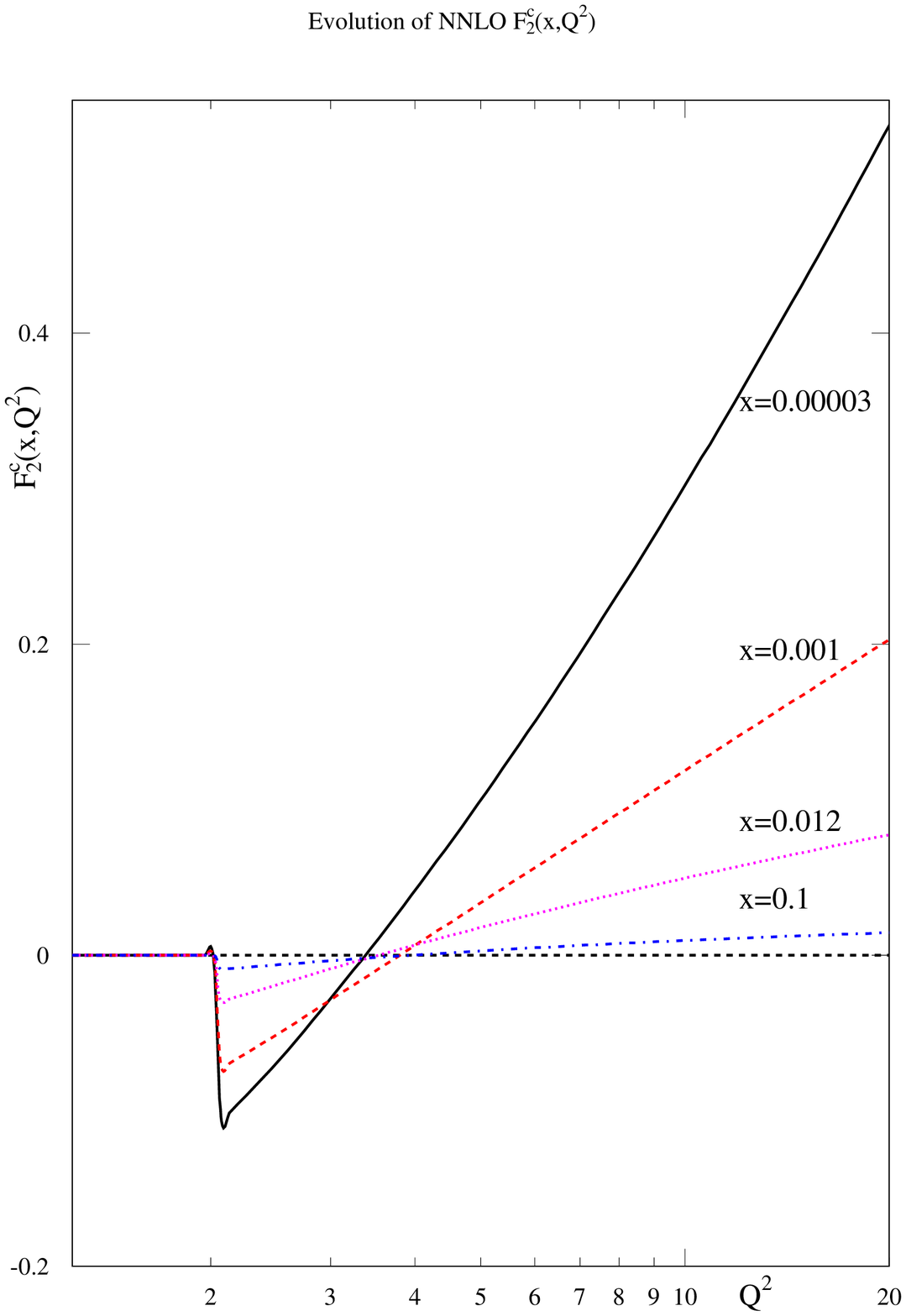}\vspace{0.5cm}
\epsfxsize=0.42\textwidth\epsfbox{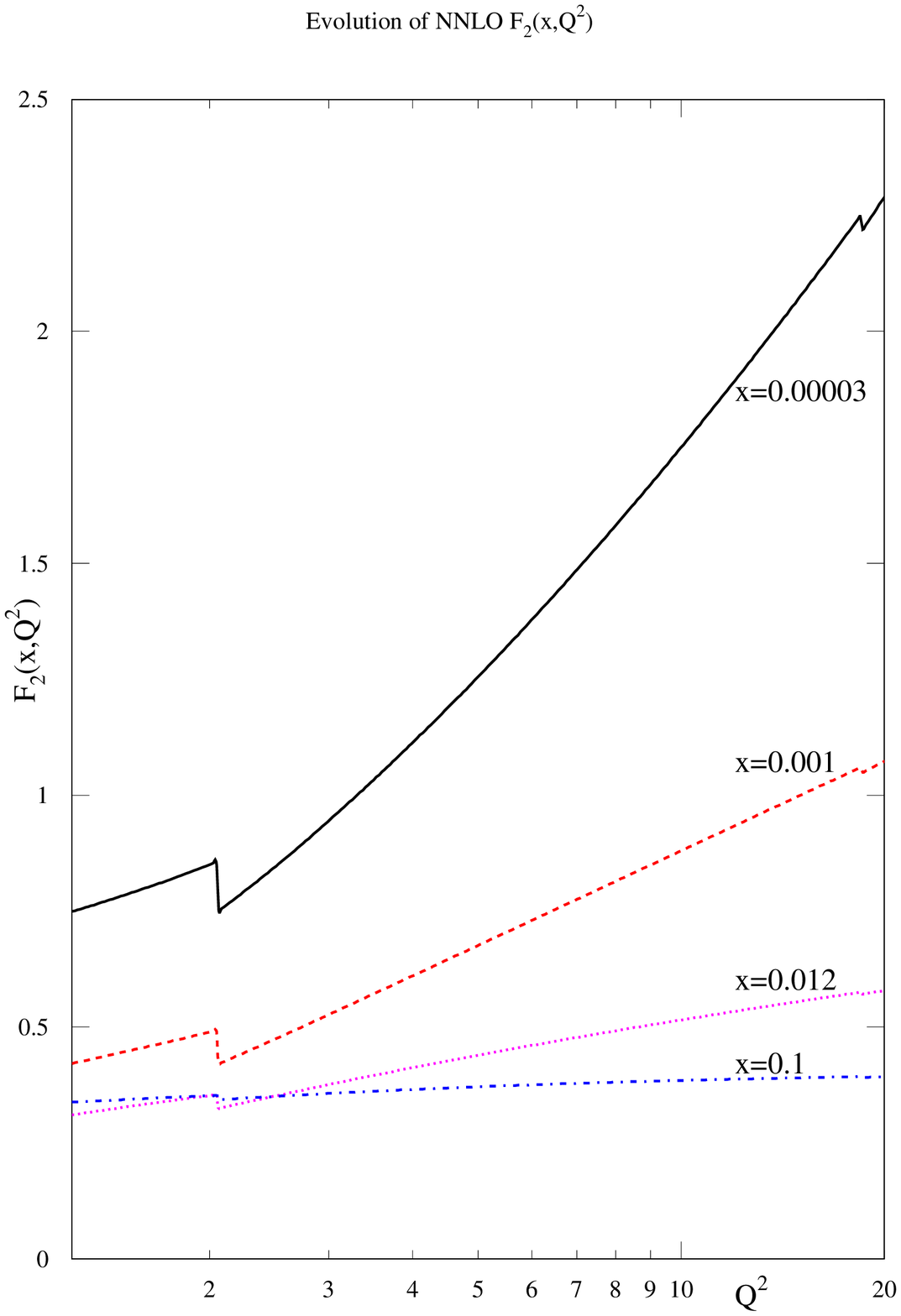}}
\caption{The discontinuity in $F_2^c(x,Q^2)$ (left) and $F_2(x,Q^2)$ (right) 
using the zero-mass variable-flavour number scheme at NNLO.}
\vspace{-1cm}
\label{fig:disc}
\end{center}
\end{figure}

At LO, i.e. zeroth order in $\alpha_S$, the relationship 
is trivial,
\beq
q(g)^{n_f+1}_k(\mu^2)\equiv q(g)^{n_f}_k(\mu^2).
\eeq
At NLO, i.e. first order in $\alpha_S$, the non-trivial contributions are  
\beq
(h+\bar h)(\mu^2)= \frac{\alpha_S}{4\pi} P^0_{qg} 
\otimes g^{n_f}(Q^2)\ln(\mu^2/m_H^2), \quad g^{n_f+1}(\mu^2)= 
\biggl(1-\frac{\alpha_S}{6\pi}\ln(\mu^2/m_H^2)\biggr)g^{n_f}(\mu^2),
\label{eq:lomatrix}
\eeq
where $h(x,\mu^2)$ is the heavy quark parton distribution.
Hence, the heavy flavour evolves from zero at $\mu^2=m_H^2$ 
according to 
standard massless quark evolution and the gluon loses 
corresponding momentum. It is natural to choose $\mu^2=m_H^2$ as the 
transition point from the $n_f$-flavour to the $n_f+1$-flavour scheme,
since at this order the partons are then continuous and the heavy quark 
starts evolving from a zero value.
At NNLO, i.e. second order in $\alpha_S$,
there is much more complication
\beq
f_{i,NNLO}^{n_f+1}(\mu^2)= \biggl(\frac{\alpha_S}{4\pi}\biggr)^2 \sum_{ij}  
(A^{2,0}_{ij}+A^{2,1}_{ij}\ln(\mu^2/m_H^2)+A^{2,2}_{ij}\ln^2(\mu^2/m_H^2))
\otimes f_j^{n_f}(\mu^2),
\eeq
where the $A^{2,0}_{ij}$ \cite{buza} are 
generally nonzero. There is no longer any 
possibility of a smooth transition at $\mu^2=m_H^2$. Since each of the 
$A^{k,0}_{ij}$ is a different function there is no smooth turn on of the 
flavour distribution at any alternative value of $\mu^2$, hence, for 
technical simplicity, it seems sensible to keep the transition point at 
$\mu^2=m_H^2$, though any other point can be chosen if desired. 
Making this choice it turns out that $A^{2,0}_{Hg}$, the matrix element
giving the gluon contribution to the heavy quark distribution, is negative at 
small $x$, even though the structure function is always positive, 
and the heavy quark starts evolving from a negative value in the 
$\msb$ scheme. This highlights the fact that parton distributions are not 
physical quantities. However, it also illustrates a major problem with
the zero-mass variable-flavour number scheme. 

In order to make a concrete illustration of the effect 
we must choose a factorization and 
renormalization scale. For light partons both of these are conventionally 
chosen to 
be $\mu^2=Q^2$. It is most natural to place the heavy flavour on the 
same footing, and choose the same scale, e.g. it is difficult to think of 
momentum conservation at a given scale if the factorization scale is 
different for different partons. Hence, for the remainder of this article 
I will set $\mu^2=Q^2$. Having made a choice, we see that    
in the ZMVFNS the coefficient functions already lead 
to a discontinuity in the structure functions at NLO, i.e.
\bea
F_2^H(x,Q^2) \qquad&=&0 \qquad Q^2 < m_H^2,\nonumber\\
&=& 
\frac{\alpha_S}{4\pi}
C_{2,g}\otimes g^{n_f+1}(Q^2) \qquad Q^2>m_H^2.
\eea
However, this is a very small effect at NLO.
It is larger at NNLO, since the coefficient function makes a larger
contribution, 
even though it is nominally of higher order, and it is  
negative at smallish $x$ ($x  \sim 0.001$), as is the starting value 
of the partons. 
Hence, $F^H_2(x,Q^2)$ is also negative for $Q^2=m_H^2$.
This shows that  the ZMVFNS is not really feasible at NNLO. It leads to huge 
discontinuities in the charm structure function $F^c_2(x,Q^2)$, and even 
significant 
discontinuities in the total structure function 
$F_2(x,Q^2)$. These are shown in Fig.\ref{fig:disc}, 
where a small discontinuity in $F_2(x,Q^2)$ at $Q^2=m_b^2$ is also seen. 
This is a measure of the mistake made in omitting the ${\cal O}(m_H^2/Q^2)$ 
corrections in this approximate ``scheme''. One really
needs a general Variable-Flavour Number Scheme (VFNS)
joining the two well-defined limits of $Q^2\leq m_H^2$ and 
$Q^2\gg m_H^2$ in a theoretically correct manner.  
We will outline such a scheme, which has been implemented explicitly up to 
NNLO, in the remainder of this article.\footnote{The scheme has 
previously been outlined
in a very brief form in \cite{dis05}.}

\section{The Variable-Flavour Number Scheme}

A correct VFNS can be defined by demanding equivalence of the 
$n_{f}$ (FFNS) and $n_f+1$-flavour descriptions at all 
orders, i.e.    
\bea
F(x,Q^2)&=&C^{FF}_k(Q^2/m_H^2,Q^2/\mu^2)\otimes f^{n_f}_k(\mu^2)= 
C^{VF}_j(Q^2/m_H^2,Q^2/\mu^2)\otimes f^{n_f+1}_j(\mu^2)\nonumber\\
&\equiv & C^{VF}_j(Q^2/m_H^2,Q^2/\mu^2)
\otimes A_{jk}(\mu^2/m_H^2)\otimes f^{n_f}_k(\mu^2).
\eea
Hence, the VFNS coefficient functions have to satisfy
\beq
 C^{FF}_k(Q^2/m_H^2,Q^2/\mu^2) = 
C^{VF}_j(Q^2/m_H^2,Q^2/\mu^2)\otimes A_{jk}(\mu^2/m_H^2),
\label{eq:general}
\eeq
at all orders. It is important to remember that the left-hand side of this 
expression is expanded in the $n_f$-flavour coupling $\alpha_{S,n_f}$, 
while the right-hand side is most naturally expanded in the $n_f+1$-flavour 
coupling $\alpha_{S,n_f+1}$. The two are related by
\beq
\alpha_{S,n_f+1}(\mu^2)=\alpha_{S,n_f}+
\frac{1}{6\pi}\ln(\mu^2/m_H^2)\,\alpha^2_{S,n_f} 
+{\cal O}(\alpha^3_{S,n_f}).\label{eq:couplings} 
\eeq
The coupling is therefore continuous up to ${\cal O}(\alpha^2_{S})$,
but at ${\cal O}(\alpha^3_{S})$ there is a small discontinuity. This 
discontinuity does not influence the VFNS up to NNLO. 

At ${\cal O}(\alpha_S)$ eq.(\ref{eq:general}) becomes, for example, for the 
structure function $F_2(x,Q^2)$
\beq
 C^{FF,1}_{2,g}(Q^2/m_H^2) = 
C^{VF,0}_{2, HH}(Q^2/m_H^2)\otimes P^0_{qg}\ln(\mu^2/m_H^2)+
C^{VF,1}_{2,g}(Q^2/m_H^2,Q^2/\mu^2),\label{eq:LO}
\eeq
The VFNS coefficient functions automatically tend to the massless limits
as $Q^2/m_H^2 \to \infty$ \cite{collins} and, if we use 
the zeroth order cross-section for 
photon-heavy quark scattering, 
\beq
C^{VF,0}_{2, HH}(Q^2/m_H^2,z)=(1+4m_H^2/Q^2)
\delta(z-1/(1+m_H^2/Q^2)),
\eeq 
this is the original ACOT scheme \cite{acot}.  

However, $C^{VF,i}_{2,HH}(Q^2/m_H^2)$ 
(we set $\mu^2=Q^2$ explicitly for simplicity) is only uniquely defined in the
massless limit $Q^2/m_H^2 \to \infty$. One can swap ${\cal O}(m_H^2/Q^2)$
terms between $C^{VF,0}_{2, HH}(Q^2/m_H^2)$ 
and $C^{VF,1}_{2,g}(Q^2/m_H^2)$ while still satisfying Eq.(\ref{eq:LO}), i.e. 
$C^{VF,0}_{2, HH}(Q^2/m_H^2)$ is not uniquely 
defined. This is true for all $C^{VF,n}_{2, HH}(Q^2/m_H^2)$. 
The original ACOT prescription removed the ambiguity by defining 
$C^{VF,i}_{2,HH}(Q^2/m_H^2)$ as the calculated coefficient function for
an incoming massive quark. However, this  
violates the physical production threshold $W^2
> 4m_H^2$ since it only needs one quark in the final state rather than a 
quark-antiquark pair. Hence, there is not a smooth transition 
at $Q^2=m_H^2$ as $n_f \to n_f+1$.    
The Thorne-Roberts 
Variable-Flavour Number Scheme
(TR-VFNS) first recognized this ambiguity in the definition of 
$C^{VF,0}_{2, HH}(Q^2/m_H^2)$ \cite{trvfns} and removed it by 
imposition of the physically motivated constraint of $(dF_2/d\ln Q^2)$
being continuous at the transition point $Q^2=m_H^2$ (in the gluon sector).
Hence, it guaranteed smoothness at $Q^2=m_H^2$, but the approach to 
$Q^2/m_H^2 \to \infty$ is a little odd -- the VFNS result overshooting the 
zero mass result before approaching it asymptotically from above. This effect 
diminishes at higher orders but  
more of a problem is the complicated form of the scheme 
-- $C^{VF,0}_{2, HH}(Q^2/m_H^2) 
\propto (P^0_{qg})^{-1}$, which is not a simple function. This 
makes the scheme very involved at higher orders and it is also not 
well suited to charged currents \cite{TRCC}.

There have been various other alternatives since this. Most recently Tung,
Kretzer and Schmidt 
have devised the ACOT($\chi$) prescription \cite{acotchi} 
which may be interpreted as 
\bea
C^{VF,0}_{2, HH}(Q^2/m_H^2,z)&=& \delta(z-Q^2/(Q^2+4m_H^2)),\nonumber\\
 \to \qquad F^{H,0}_2(x,Q^2)&=&(h+\bar h)(x/x_{max}, Q^2), \qquad x_{max}=
Q^2/(Q^2+4m_H^2).
\eea 
Hence, the zeroth-order coefficient function tends to the standard 
$C^{ZM,0}_{2, HH}(z)= \delta(1-z)$
for $Q^2/m_H^2 \to \infty$ but respects the threshold requirement $W^2
=Q^2(1-x)/x \ge 4m_H^2$ for quark-antiquark production. Moreover, it is 
very simple.  For the VFNS to remain simple (and physically motivated) at all 
orders $n$ in $\alpha_S$  it is necessary to choose
\beq
C^{VF,n}_{2, HH}(Q^2/m_H^2,z)= C^{ZM,n}_{2, HH}(z/x_{max}).
\label{eq:twodef}
\eeq
It is also important to choose
\beq
C^{VF,n}_{L, HH}(Q^2/m_H^2,z)\propto C^{ZM,n}_{L, HH}(z/x_{max}),
\eeq
and to impose the condition that $C^{VF,0}_{L, HH}(Q^2/m_H^2,z)\equiv 0$
(as is done in \cite{trvfns}), 
despite the fact that one obtains 
\beq
C^{0}_{L, HH}(Q^2/m_H^2,z) = 4z\frac{m_H^2}{Q^2}
\delta(z-1/(1+m_H^2/Q^2))
\eeq
for single quark-photon scattering. 
Because the heavy-flavour contribution to $F_L(x,Q^2)$ is highly suppressed 
at low values of $Q^2$, compared to that for $F_2(x,Q^2)$, we choose
\beq
C^{VF,n}_{L, HH}(Q^2/m_H^2,z)=\frac{5}{4}\biggl(\frac{1}{1+4m_H^2/Q^2}
-\frac{1}{5}\biggr) 
C^{ZM,n}_{L, HH}(z/x_{max}).
\label{eq:Ldef}
\eeq
The prefactor is zero at $Q^2=m_H^2$ and tends to unity as $Q^2/m_H^2 \to
\infty$, but is suppressed by the physical threshold of $4m_H^2$ for
intermediate values of $Q^2$. This factor guarantees 
that the heavy-flavour parton contribution to $F_L^H(x,Q^2)$ is 
heavily moderated for $Q^2$ just above $m_H^2$. Failure to do this results in 
a kink in $F_L^H(x,Q^2)$ just above $Q^2=m_H^2$. The fact that this is a 
function of $Q^2/m_H^2$ only avoids any additional complications when 
obtaining higher-order coefficient functions by convolutions with matrix 
elements which the definition in Eq.(\ref{eq:general}) requires.

Adopting this convention for the heavy-flavour coefficient functions at 
NNLO we have, for example, 
\bea
C^{VF,2}_{2, Hg}(Q^2/m_H^2,z)&=& C^{FF,2}_{2, Hg}(Q^2/m_H^2,z) - 
C^{ZM,1}_{2, HH}(z/x_{max})\otimes
A^1_{Hg}(Q^2/m_H^2)\nonumber\\
& &\hspace{-3.8cm}-C^{ZM,0}_{2, HH}(z/x_{max})\otimes A^2_{Hg}(Q^2/m_H^2)
-\frac{1}{6\pi}\ln(Q^2/m_H^2)\,C^{ZM,0}_{2, HH}(z/x_{max})
\otimes A^1_{Hg}(Q^2/m_H^2).\label{eq:nnlodisc}
\eea
The last term comes from the change in the coupling constant as we go 
across the transition point, i.e. from Eq.(\ref{eq:couplings}).
This would be absent if we used (somewhat unnaturally) 
the $n_f$-flavour renormalization scheme above $Q^2=m_H^2$, as is sometimes 
done, but this means the definition of $A^2_{Hg}(Q^2/m_H^2,z)$  
is different in the two renormalization schemes (compare that in 
\cite{buza} with that in \cite{buza2}).  
There is also in principle a contribution of the form 
\beq
-\frac{1}{6\pi}\ln(Q^2/m_H^2)\,C^{VF,1}_{2, Hg}(Q^2/m_H^2,z)-
C^{VF,1}_{2, Hg}(Q^2/m_H^2,z)\otimes A^1_{gg}(Q^2/m_H^2)
\eeq
on the right-hand side, but 
\beq
A^1_{gg}(Q^2/m_H^2,z)= 
-\frac{1}{6\pi}\ln(Q^2/m_H^2)\,\delta(1-z), 
\eeq
as seen in Eq.(\ref{eq:lomatrix}), so these terms cancel.
Both would be absent if we used the $n_f$-flavour renormalization 
scheme above $Q^2=m_H^2$.

From the definition in Eq.(\ref{eq:nnlodisc}) we see that  
since $A^2_{Hg}(1,z)\not=0$, 
the coefficient function 
$C^{2}_{2, Hg}(Q^2/m_H^2,z)$
is discontinuous as we go across $Q^2=m_H^2$. This compensates exactly for 
the ${\cal O}(\alpha_S^2)$ 
discontinuity arising from that in the heavy-flavour parton distribution, i.e.
for the term $C^{VF,0}_{2, HH}\otimes (h+\bar h)$, and  
$F^H_2(x,Q^2)$ is continuous.\footnote{In principle there are 
${\cal O}(\alpha_S^3)$ discontinuities
due to terms such as $C^{VF,1}_{2, HH}\otimes (h+\bar h)$ and 
$C^{VF,1}_{2, Hg}\otimes g^{n_f+1}$, i.e. ${\cal O}(\alpha_S)$ coefficient 
functions convoluted with ${\cal O}(\alpha_S^2)$ discontinuities in partons. 
These would be cancelled at NNNLO by discontinuities in 
${\cal O}(\alpha_S^3)$ coefficient functions, but are actually tiny 
in practice.}  
In practice this requires the knowledge of $C^{FF,2}_{2, Hg}(Q^2/m_H^2,z)$.
An expression for this exists as semi-analytic code \cite{code} where
the dominant contributions for 
$W^2 \to \infty$ and $W^2 \to 4m_H^2$ (from above) are 
analytic and the rest numerical.  
I have produced much 
faster analytic expressions which are exact for $Q^2/m_H^2 \to \infty$ 
and in some cases for $W^2 \to 4m_H^2$, and 
the $(m_H^2/Q^2)$ remainders are provided by fitting the values to 
analytic functions with a number of free parameters. These final expressions
are slightly approximate,
but the error in $F_2^H(x,Q^2)$ is only $\sim 1\%$ even in the most extreme 
cases. 

\begin{figure}
\begin{center}
\centerline{\epsfxsize=0.7\textwidth\epsfbox{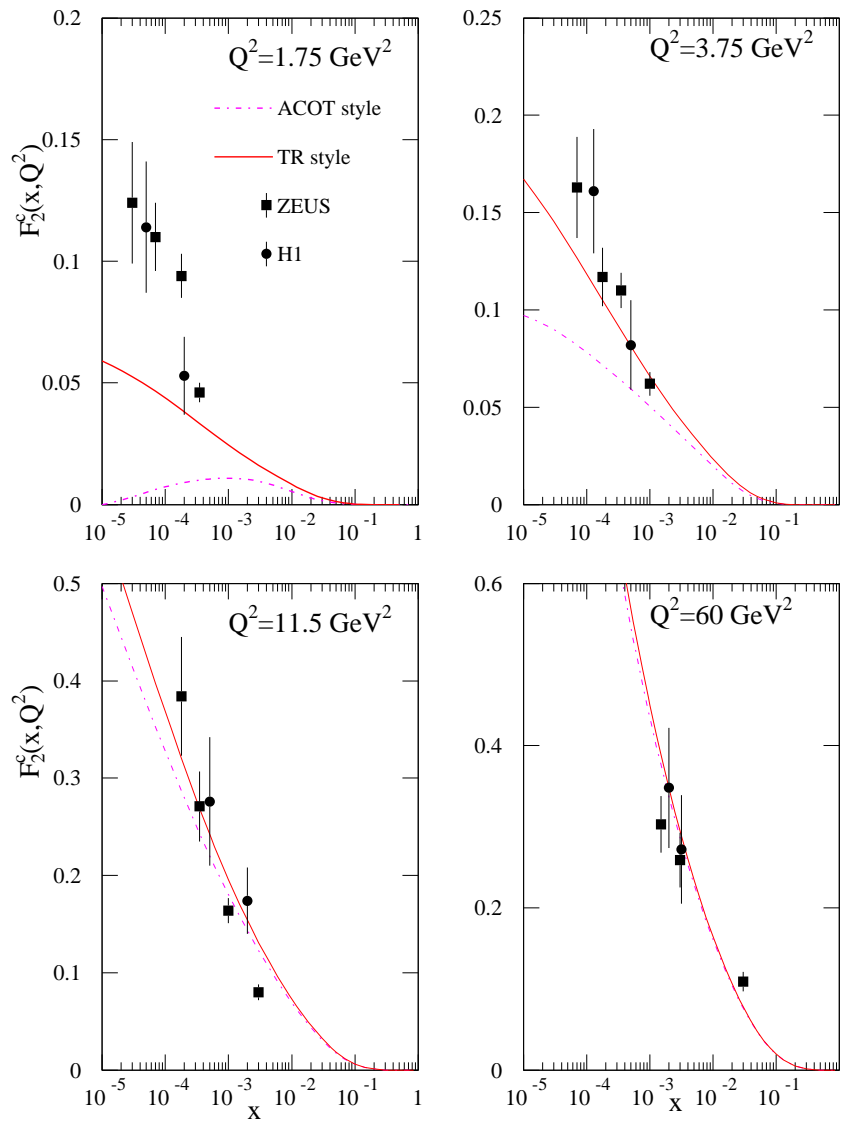}}
\caption{Comparison between the ACOT choice of ordering and the 
Thorne-Roberts choice at NLO.}
\vspace{-1cm}
\label{fig:order}
\end{center}
\end{figure}

There is one more problem in defining the VFNS. The ordering for 
$F_2^H(x,Q^2)$ is different for the $n_f$ and $n_f+1$ regions. This can be 
illustrated by the following table which shows the order by order expressions 
both below and above the transition point. 
\vspace{0.2cm}
$$
\begin{array}{ccc}
&n_f-{\rm flavour}& n_f+1-{\rm flavour} \\
&&\\
LO &\frac{\alpha_S}{4\pi} 
C^{FF,1}_{2, Hg}\otimes g^{n_f} &  
C^{VF,0}_{2, HH}\otimes (h+\bar h)\\
NLO &\biggl(\frac{\alpha_S}{4\pi}\biggr)^2
\!\!(C^{FF,2}_{2, Hg}\otimes g^{n_f}\!+\!C^{FF,2}_{2, Hq}\otimes \Sigma^{n_f})
&\frac{\alpha_S}{4\pi}(C^{VF,1}_{2, HH}\otimes (h+\bar h)
+C^{VF,1}_{2, Hg}\otimes g^{n_{f+1}})\\
NNLO &\biggl(\frac{\alpha_S}{4\pi}\biggr)^3
\sum_i C^{FF,3}_{2, Hi}\otimes f_i^{n_f} &
\biggl(\frac{\alpha_S}{4\pi}\biggr)^2\sum_j
C^{VF,2}_{2, Hj}\otimes f^{n_f+1}_j.
\end{array}
$$
The issue is that the series expansion begins at zeroth order above the 
transition point, where there is a heavy flavour distribution, but at 
${\cal O}(\alpha_S)$ below the transition point. Hence, what is meant by LO, 
NLO {\it etc.} is different by one power of $\alpha_S$ as one changes the 
number of active quark flavours. Therefore, making the transition directly 
from a given fixed order to the same relative order when going from 
$n_f$ to $n_f+1$ flavours leads to a different order in $\alpha_S$
and discontinuities which may be rather 
significant-- for example, LO is nonzero as one approaches the transition 
point from below, but zero when approaching it from above. One 
must make some decision on how to deal with this problem.

Up to now ACOT have used the same order of $\alpha_S$ above and below the 
transition point,  e.g. at NLO
\beq
\frac{\alpha_S}{4\pi} 
C^{FF,1}_{2, Hg}\otimes g^{n_f} \to C^{VF,0}_{2, HH}\otimes (h+\bar h)
+ \frac{\alpha_S}{4\pi}
(C^{VF,1}_{2, HH}\otimes (h+\bar h)
+C^{FF,1}_{2, Hg}\otimes g^{n_f+1}).
\eeq 
The structure function is then automatically continuous. However, 
there is effectively LO evolution below the transition point -- 
$C^{FF,1}_{2, Hg}$ contains only information on $P^0_{qg}$, 
not on $P^1_{qg}$ --
and NLO evolution above it. Hence the slope $d\,F_2^H(x,Q^2)/d\,\ln Q^2$ is 
discontinuous.

The Thorne-Roberts scheme used the same relative order above and below 
the transition point, but added a uniquely determined
$Q^2$-independent term above the transition point to 
maintain continuity of the structure function. For example, at LO 
\beq 
\frac{\alpha_S(Q^2)}{4\pi} 
C^{FF,1}_{2, Hg}(Q^2/m_H^2)\otimes g^{n_f}(Q^2) \to \frac{\alpha_S(M^2)}{4\pi} 
C^{FF,1}_{2, Hg}(1)\otimes 
g^{n_f}(M^2)+ C^{VF,0}_{2, HH}(Q^2/m_H^2)\otimes (
h+\bar h)(Q^2),
\eeq
i.e. this prescription 
freezes the higher order $\alpha_S$ term when going upwards 
through $Q^2=m_H^2$.
This difference in choice is extremely important at low $Q^2$
(if using $\mu^2 =Q^2$), as is illustrated in Fig.\ref{fig:order} which 
compares the two choices at 
NLO. The ${\cal O}(\alpha_S^2)$ part is dominant at low $x$ for 
$Q^2\leq m_c^2$ because the ${\cal O}(\alpha_S^2)$ coefficient functions 
diverge at small $x$ whereas the ${\cal O}(\alpha_S)$ coefficient function 
is finite in this limit. Indeed, the ``frozen'' part is very significant for 
$m_c^2 \leq Q^2 \leq\,12\GeV^2$. Its inclusion clearly improves the match 
to the data \cite{zeus,h1}. It is also clear that
switching from the standard $n_f$-flavour NLO to the 
standard $n_f+1$-flavour NLO would lead to a large discontinuity
in $F^H_2(x,Q^2)$. Hence we choose to continue using the  
Thorne-Roberts approach for the matching at the transition point. 
However, this is a place where there is a definite freedom of choice, and 
there are various possibilities available.

\begin{figure}
\begin{center}
\centerline{\epsfxsize=0.7\textwidth\epsfbox{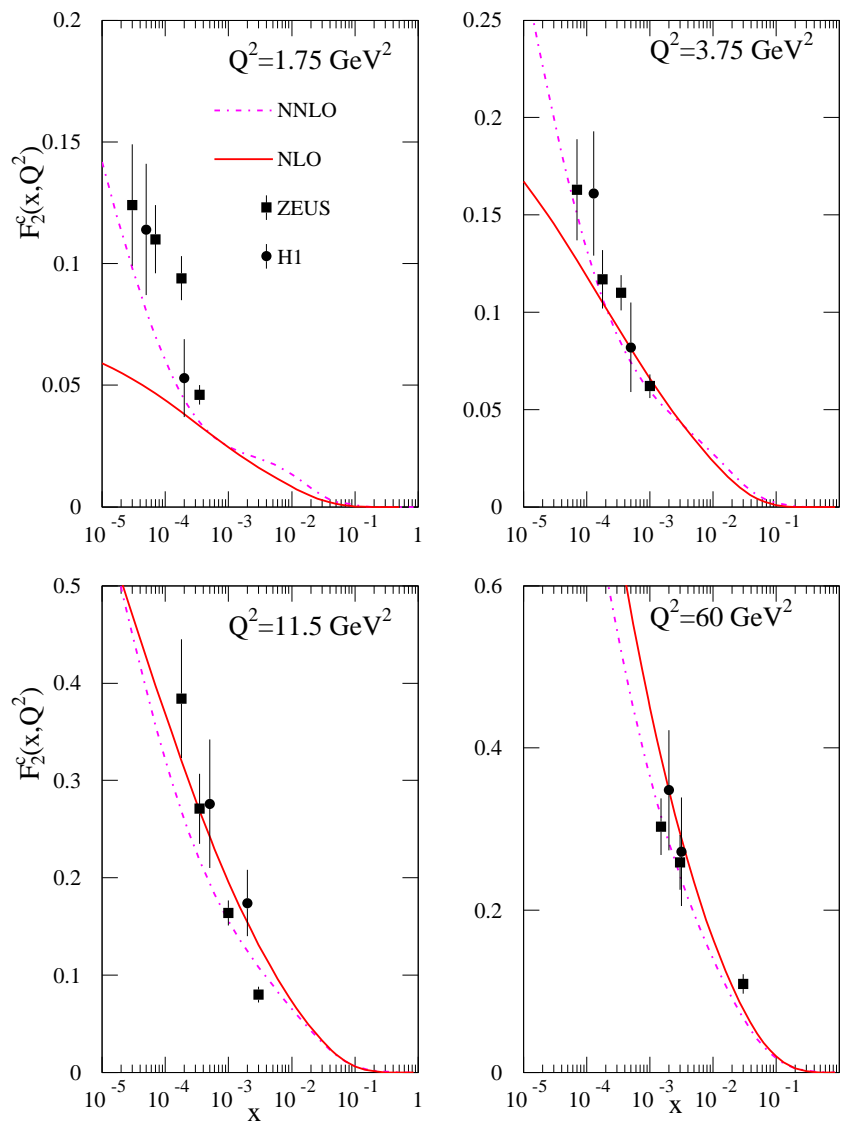}}
\caption{Comparison of the NLO and NNLO 
predictions for $F_2^c(x,Q^2)$ compared 
with data.}
\vspace{-1cm}
\label{fig:comp}
\end{center}
\end{figure}

With the type of 
choice made for the definition of the heavy flavour coefficient 
functions $C_{L,HH}^{VF,n}(Q^2/m_H^2,z)$ in Eq.(\ref{eq:Ldef}) there
is no problem with ordering across the transition point for the longitudinal 
structure function. This is because both the FFNS and VFNS coefficient 
functions begin at ${\cal O}(\alpha_S)$ for both the gluon and heavy quarks. 
It is possible to choose a non-vanishing value for 
$C_{L,HH}^{VF,0}(Q^2/m_H^2,z)$, and indeed some VFNS definitions do so, 
but this is contrary to the spirit of this approach, and will lead to extra
complications.

In order to define fully the VFNS at NNLO, this choice for ordering above 
and below the transition point means that we need the 
${\cal O}(\alpha_S^3)$ heavy-flavour coefficient functions for 
$Q^2 \leq m_H^2$ and that the contribution from these should be 
frozen for $Q^2>m_H^2$. However, these coefficient functions are not 
yet known. Nevertheless,  
we do know the leading threshold logarithms \cite{thresh}, 
i.e. the leading contribution for $W^2$ not much above $4m_H^2$. This is 
given by 
\beq
C^{FF,3, thresh}_{2, Hg}(Q^2/m_H^2,z) \sim \frac{1}{z(1+\eta)}
\frac{Q^2}{Q^2+4m_H^2}
\rho(\eta,Q^2/m_H^2), \qquad \eta= \frac{Q^2(1-z)}{z4m_H^2}-1,
\label{eq:thresh}
\eeq
i.e. $\eta \to 0$ at threshold and $\eta \to \infty$ as
$W^2 \to \infty$. $\rho(\eta,Q^2/m_H^2)$ is a function which 
models the contribution from the dominant 
threshold logarithms. This contribution occurs only in the gluon sector.

\begin{figure}
\begin{center}
\centerline{\epsfxsize=0.6\textwidth\epsfbox{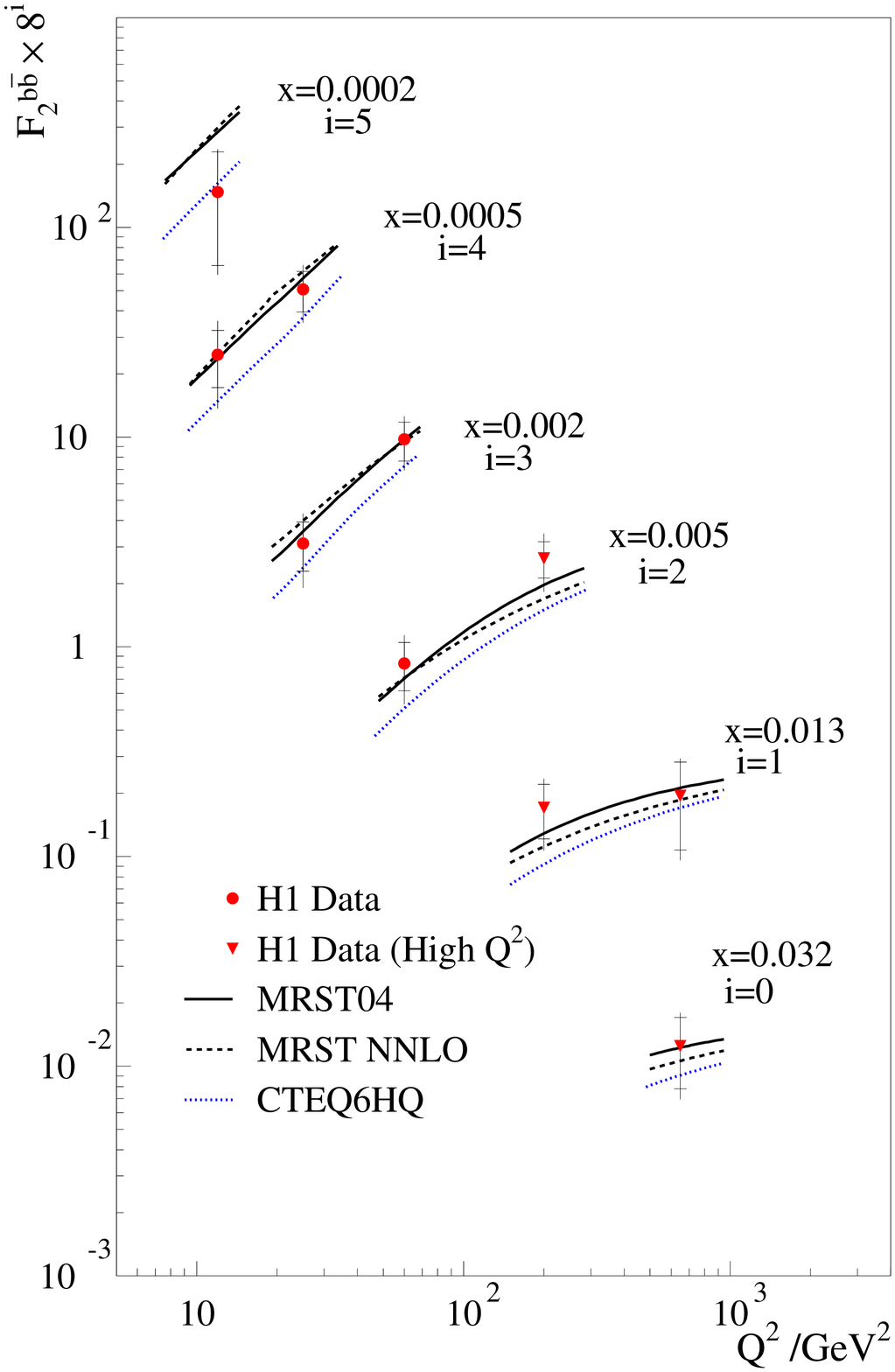}}
\caption{Comparison of the NLO and 
NNLO predictions for $F_b^c(x,Q^2)$ compared 
with H1 data.}
\vspace{-1cm}
\label{fig:bottom}
\end{center}
\end{figure}

We can also derive the leading $ln(1/x)$ term from $k_T$-dependent
impact factors derived by Catani, Ciafaloni and Hautmann \cite{asymp}.
With a little work these can be shown to give
\beq
C^{FF,3, low x}_{2, Hg}(Q^2/m_H^2,z) = 96\frac{\ln(1/z)}{z}
\kappa_2(Q^2/m_H^2), 
\eeq
where $\kappa_2(Q^2/m_H^2)$ may be calculated and $\kappa_2(1) 
\approx 4$.
We also know that in this small-$x$ limit 
$C^{FF,3, low x}_{2, Hq}(Q^2/m_H^2,z)=
4/9\,C^{FF,3, low x}_{2, Hg}(Q^2/m_H^2,z)$. 
By analogy with the known NNLO coefficient functions 
and splitting functions it is reasonable to propose that this be modified to 
\beq
C^{FF,3, low x}_{2, Hg}(Q^2/m_H^2,z) = \frac{96}{z}\beta(\ln(1/z)-4)
(1-z/x_{max})^{20} \kappa_2(Q^2/m_H^2).
\label{eq:asymp2}
\eeq
$\beta=(1-4m^2_Hz/(Q^2(1-z)))^{1/2}$ is the velocity of a heavy quark in the 
centre-of-mass frame and its introduction
ensures that this contribution $\to 0$ smoothly at threshold.
The leading $\ln(1/z)$ is accompanied by $\sim -4$, i.e. a $1/z$ term of 
similar size to that in other known coefficient 
functions and splitting functions 
is introduced. Finally, the effect of the 
this entire small $z$ term is damped as $z \to 1$ by the large power of 
$(1-z/x_{max})$. The power of $20$ is chosen in order to make the 
contribution in Eq.(\ref{eq:asymp2}) very suppressed until $x<0.1$, which
is in line with the values of $x$ above which the small-$x$ divergent 
terms tend to be suppressed in the complete NNLO splitting 
functions \cite{NNLOs}.     
The total approximate NNLO coefficient function is obtained by adding the 
contributions in Eq.(\ref{eq:thresh}) and Eq.(\ref{eq:asymp2})
The amount of information is similar to (though a little weaker than) 
that used previously to derive
approximate NNLO splitting functions \cite{approx}, which turned out to
be a very good approximation once the exact expressions became known.

These expressions could also be used to provide an approximate
NNLO FFNS definition. However, in this case they would have to be used over 
a far wider range of $Q^2$, rather than the small range here. In particular 
the frozen NNLO contribution becomes a proportionally very small
contribution to the total $F_2^H(x,Q^2)$ at high $Q^2$, whereas there is no 
reason to believe the ${\cal O}(\alpha_S^3)$ contribution in the FFNS is small
since it contains terms of order $\ln^3(Q^2/m_H^2)$. The NNLO FFNS would 
therefore be more genuinely approximate. For $F_L^H(x,Q^2)$ we can approximate
the ${\cal O}(\alpha_S^3)$ coefficient functions at $Q^2\leq m_H^2$ using the 
same approach. In this case there is no large threshold contribution and 
the small-$x$ contribution is estimated to be  
\beq
C^{FF,3, low x}_{L, Hg}(Q^2/m_H^2,z) = \frac{96}{z}\beta^3(\ln(1/z)-4)
(1-z/x_{max})^{20}  \kappa_L(Q^2/m_H^2),
\label{eq:asympL}
\eeq
where $\kappa_L(1) \approx 0.16$ and the $\beta^3$ 
reflects the fact that the longitudinal heavy-flavour coefficients are much 
more suppressed near threshold. For the VFNS for $F^H_L(x,Q^2)$ one can then 
extrapolate smoothly to the exact massless 
${\cal O}(\alpha_S^3)$ coefficient functions used in the VFNS at high $Q^2$.
Explicitly, for $Q^2> m_H^2$
\bea
F_L^H(x,Q^2)&=& \frac{5}{4}\biggl(1-\frac{1}{1+4m_H^2/Q^2}\biggr)\Biggl(
\biggl(\frac{\alpha_S}{4\pi}\biggr)^3
\sum_i C^{FF,3}_{L, Hi}\otimes f_i^{n_f}-C^{FF,1}_{L, Hg}\otimes
A^2_{gg,H}\otimes g^{n_f}\Biggr)_{Q^2=m_H^2}\nonumber\\
&&\hspace{0.8cm}+\frac{5}{4}\biggl(\frac{1}{1+4m_H^2/Q^2}-\frac{1}{5}\biggr)
\biggl(\frac{\alpha_S(Q^2)}{4\pi}\biggr)^3
\sum_i C^{ZM,3}_{L, Hi}\otimes f_i^{n_f+1}(Q^2),
\eea
where the term $\propto C^{FF,1}_{L, Hg}\otimes
A^2_{gg,H}\otimes g^{n_f}$ maintains continuity of the structure function 
across the transition point despite the discontinuity in the gluon 
distribution. 
Again the degree of modelling and approximation is far less than in a
NNLO FFNS.

Using these approximate ${\cal O}(\alpha_S^3)$ coefficient functions one 
can produce full NNLO predictions for structure functions with 
discontinuous partons and coefficient functions but continuous $F^H(x,Q^2)$. 
The results are not very sensitive to the choices made in this
approximation, as long as they are within a sensible 
range. Note also that the definition of the VFNS, relying only on 
Eq.(\ref{eq:general}), Eq.(\ref{eq:twodef}), Eq.(\ref{eq:Ldef}) and the 
ordering across the transition point, may be straightforwardly generalised to
any choice of factorization and renormalization scales. For the simplest 
choice of $\mu^2=Q^2$ the NNLO corrections are seen in Fig.\ref{fig:comp}. 
They clearly improve the match to the lowest $Q^2$ data, where NLO 
is always too low. This large increase at low $x$ is because the NNLO 
coefficient functions are more divergent than the NLO coefficient functions. 
The comparison with the recent bottom quark production 
data \cite{bottom} from H1 is also 
shown in Fig.\ref{fig:bottom}.\footnote{This is an updated version of Fig.8
in \cite{bottom}. This previous figure used the MRST04 NNLO partons which 
applied a rather approximate NNLO treatment of heavy flavours. The updated 
figure is constructed using NNLO partons from a fit which applies the 
NNLO VFNS. This leads to 
a generally slightly increased prediction for $F^b_2(x,Q^2)$.
The gluons from this new fit are exhibited in Fig.3 of \cite{Ringberg}, 
and are more negative than our previous, approximate NNLO partons at 
low $Q^2$ and very small $x$.}  
The agreement is good, but one can clearly 
see that the slope $(d F_2^b(x,Q^2)/d\ln Q^2)$ is smaller at NNLO than at 
NLO, and the same is true for charm. This is because one is interpolating 
from a higher value at low $Q^2$,
due to the NNLO coefficient function contribution, to lower values at high
$Q^2$, where the quark content is dominant and has a relative reduction 
compared to NLO because the evolution has started from a negative value rather 
than zero. Hence, this tendency for a reduction in the slope of the 
heavy-flavour structure function at NNLO is a generic feature of NNLO. 
The detailed phenomenology of the global fit with 
the NNLO VFNS prescription will appear in the account our next global parton 
analysis \cite{MRST06}.

\section{Light-Flavour Sector}

At NNLO it is no longer possible to think of heavy-flavour effects as only
associated with heavy flavour production, as has already been discussed in 
\cite{smith}. At this order we 
also get contributions due to heavy flavours away from 
the photon vertex. Examples of such contributions are shown below.  

\vspace{-0.1cm} 

\begin{center}
\begin{picture}(450,200)(0,0)
\SetColor{Red}
\Photon(10,190)(40,160){5}{6}
\Text(5,195)[]{$\gamma^{\star}$}
\ArrowLine(40,10)(40,160)
\ArrowLine(40,160)(190,160)
\Gluon(40,40)(80,80){5}{2}
\Gluon(120,120)(160,160){5}{2}
\ArrowArc(100,100)(28.28,215,45)
\ArrowArc(100,100)(28.28,45,215)
\Text(72,128)[]{$\bar h$}
\Text(128,72)[]{$h$}
\Text(40,3)[]{$q$}
\Text(197,160)[]{$q$}
\Text(225,100)[]{$+$}
\Photon(260,190)(290,160){5}{6}
\Text(255,195)[]{$\gamma^{\star}$}
\ArrowLine(290,10)(290,160)
\ArrowLine(290,160)(440,160)
\Gluon(290,60)(390,60){7}{4}
\ArrowLine(390,60)(440,110)
\ArrowLine(440,10)(390,60)
\Text(447,110)[]{$h$}
\Text(447,10)[]{$\bar h$}
\Text(290,3)[]{$q$}
\Text(447,160)[]{$q$}
\end{picture}
\end{center}
 
\vspace{-0.1cm}

In the light-flavour sector the VFNS is defined as before. There are
matrix elements changing the light flavours as we go from $n_f$ flavours to
$n_f+1$ flavours and contributions to light-flavour coefficient functions 
in the FFNS from the type of diagrams above. The exact expressions for 
these, as well as the asymptotic limits, can be found in \cite{buza2}. 
Defining the consistent VFNS according to Eq.(\ref{eq:general}) 
then leads to a discontinuity in 
the coefficient functions across the transition point,
which again cancels that in the light quark distributions, leaving the total
structure function continuous. 
In this sector there are peculiar complications due to the occurrence of 
$(\ln^m(1-z)/(1-z))_+$ terms at the threshold. For example, from the type of 
diagrams on the right-hand side we get contributions to the structure 
function of the form
\bea 
\alpha_S^2\ln^2(Q^2/m_H^2)\biggl(\frac{ln^m(1-z)}{1-z}\biggr)\otimes 
q(x,Q^2) &=&
\alpha_S^2\ln^3(Q^2/m_H^2)q(x,Q^2)\nonumber \\
&& \hspace{-4.5cm}+ 
\alpha_S^2\ln^2(Q^2/m_H^2)\biggl(\frac{\ln^m(1-z)}{1-z}\biggr)_
+\otimes q(x,Q^2)
+{\cal O}(\ln^2(Q^2/m_H^2)).
\eea
The maximum value of $z$ in the convolution is $1/(1+4m_H^2/Q^2)$ so the 
divergence at $z=1$ is not reached. However, the convolution on the left-hand 
side does produce an additional power of $\ln(Q^2/m_H^2)$. 
The $\ln^3(Q^2/m_H^2)$ contributions then cancel exactly with contributions 
from terms of the form $\alpha_s^2\ln^3(Q^2/m_H^2)\delta(1-z)$ in the 
coefficient functions from the type of diagrams on the 
left-hand side, leaving the remainder, including the 
``$+$''-distributions. 
This is an added complication compared to the case where 
the heavy quark appears at the photon vertex, requiring particular 
care in a numerical implementation of the VFNS. However, it is only really 
a technical problem rather than producing any fundamentally new features.
In the light-flavour
sector there is no problem with ordering. The light-flavour contribution 
to $F_2(x,Q^2)$ begins at zeroth order in the FFNS and VFNS.

Furthermore, if one is being totally correct, the left-hand type 
diagram and the soft parts of the right-hand type 
diagram should contribute to the light-flavour structure function,
and the hard part of the right-hand type 
diagram contributes to $F^H_2(x,Q^2)$ \cite{smith}. 
This can be implemented (it depends on a separation parameter, determining
``hard'' and ``soft''), but each 
contribution is in practice tiny. At the moment we include all such
contributions in the light flavours. This leads to a very small 
underestimate of the heavy-flavour structure functions.  

\section{Charged-Current Structure Functions}

The VFNS works, in principle, in much the same way for charged currents. 
The zeroth-order coefficient function for single (anti)charm production
from a (anti)strange quark is now  
\beq
C^{VF,0}_{2, sc}(Q^2/m_c^2,z)= \delta(z-Q^2/(Q^2+m_c^2)),
\eeq
i.e. the threshold is now for a single heavy quark production. The same is 
obviously true for (anti)charm production from a (anti)down quark. This
is far simpler than the prescription in \cite{TRCC}. At higher orders the 
generalization is again trivial. Using the ACOT($\chi$) reasoning one simply
uses the light coefficient function with argument $z$ replaced by 
$z/(1+m^2_c/Q^2)$. However, there is a major problem in defining the 
full VFNS, even at NLO. The FFNS coefficient functions are 
calculated at ${\cal O}(\alpha_S)$ \cite{gkr} but are not yet calculated
exactly at ${\cal O}(\alpha_S^2)$ for the charged-current case.\footnote{They 
have been determined in the $Q^2 \gg m_H^2$ limit in \cite{nlocc}.} 
The generalization from the neutral 
current case is not simple because the divergences in the final state are
different, i.e. one more particle is massless and regularized by 
dimensional regularization, and one fewer is regularized by the mass.

Hence, in defining the VFNS for the charged current case it is
necessary to 
make an approximation. In practice we have used the ${\cal O}(\alpha_S^2)$
neutral current cross-sections but altered the threshold dependence in all 
expressions so that all dependence on $Q^2/(Q^2+4m_H^2)$ is replaced by 
dependence on $Q^2/(Q^2+m_H^2)$. This guarantees that all terms respect the 
true kinematic threshold. The approximation occurs mainly at low $Q^2$ where 
the ${\cal O}(\alpha_S^2)$ FFNS coefficient functions are most important in 
the VFNS. This is not peculiar to this definition 
of a VFNS, but will be present in any current 
approach.\footnote{This is except for the ZMVFNS, which will simply be 
completely wrong by terms of ${\cal O}(m_c^2/Q^2)$ and will have 
incorrect kinematic thresholds, whereas the approximation here is guaranteed 
to have the correct general form.}
There is an uncertainty at low $Q^2$ at ${\cal O}(\alpha_S^2)$ that can only
be removed by an explicit calculation. At higher $Q^2$ the VFNS tends 
to the zero-mass limit for all 
coefficient functions, so all expressions become exact. Hence the HERA charged
current data are very insensitive indeed to the approximation. The charged 
current data at low $Q^2$ only exist down to $x\sim 0.01$, much higher than 
for the neutral current HERA data, and at low $Q^2$ and high-$x$ heavy-flavour 
production is very small. Hence any errors in the approximation are not very
important phenomenologically.   

\section{Conclusions}

There are discontinuities in both the parton distributions and the coefficient 
functions at NNLO. This makes a variable-flavour number scheme 
more necessary than ever. The ZMVFNS is badly discontinuous
at the transition point $Q^2=m_H^2$, and the FFNS
is only approximate at NNLO. A generalization of the ACOT($\chi$)
prescription leads to a physically sensible and simple VFNS,
in principle defined to all orders. 
One must still be careful about matching when going across the transition 
point of $Q^2=m_H^2$. If this matching is done properly it 
guarantees the continuity of the physical structure functions
and maximises the smoothness of the function. 
We choose the Thorne-Roberts 
method of matching above and below the transition, 
i.e. choose the correct order for the region of $n_f$ flavours 
and add an additional, uniquely defined constant for the region of $n_f+1$
flavours to guarantee continuity. 
This choice is significant and leads to a much better match to 
the low $Q^2$ data. 
We have devised an explicit, full NNLO VFNS for $F_2(x,Q^2)$ and $F_L(x,Q^2)$, 
with a small amount of necessary 
modelling of NNLO fixed-flavour coefficient functions. 
The NNLO variable-flavour number scheme seems to improve the fit to lowest 
$x$ and $Q^2$ data greatly and is not very sensitive to this modelling.
It is essential to use such an NNLO VFNS in NNLO global analysis of data,
and indeed the construction of the NNLO VFNS makes such a precise analysis 
possible \cite{MRST06}.

\section*{Acknowledgements}

I would like to thank Stefan Kretzer, Alan Martin, Fred Olness,  James 
Stirling, Wu-Ki Tung and Chris White 
for various discussions on this subject. I would 
also like to thank Mandy Cooper-Sarkar, Claire Gwenlan and Paul Thompson for 
numerous discussions on the comparison to data, and the latter for providing 
me with Fig.\ref{fig:bottom}. I would also like to thank
the Royal Society for the award of a University Research Fellowship. \\

\vspace{-0.6cm}


\end{document}